# Prosody of speech production in latent post-stroke aphasia

*Cong Zhang[1], Tong Li[2], Gayle DeDe[3], and Christos Salis[1]*

[1]Speech & Language Sciences, Newcastle University, UK
[2]School of Foreign Languages, Tianjin University, China
[3]Department of Communication Sciences and Disorders, Temple University, USA

`cong.zhang@newcastle.ac.uk, litong_lt@tju.edu.cn, gayle.dede@temple.edu,`
`christos.salis@newcastle.ac.uk`

## Abstract

This study explores prosodic production in latent aphasia, a mild form of aphasia associated with left-hemisphere brain damage (e.g. stroke). Unlike prior research on moderate to severe aphasia, we investigated latent aphasia, which can seem to have very similar speech production with neurotypical speech. We analysed the f0, intensity and duration of utterance-initial and utterance-final words of ten speakers with latent aphasia and ten matching controls. Regression models were fitted to improve our understanding of this understudied type of very mild aphasia. The results highlighted varying degrees of differences in all three prosodic measures between groups. We also investigated the diagnostic classification of latent aphasia versus neurotypical control using random forest, aiming to build a fast and reliable tool to assist with the identification of latent aphasia. The random forest analysis also reinforced the significance of prosodic features in distinguishing latent aphasia.

## 1. Introduction

### 1.1. Aphasia: disorder and severity

Aphasia is a language disorder which affects spoken word production as well as higher linguistic levels such as sentences and discourse. Delays in word retrieval and errors of phonological, semantic or mixed origin may also occur [1]. Problems with spoken comprehension are also common. Aphasia is associated with left-hemisphere brain damage, typically a stroke, as well as other causes, e.g., trauma or progressive conditions, when language areas are implicated. About a third of stroke survivors will develop chronic aphasia which affects people's quality of life often resulting in social isolation and poor mental health [2].

Regarding clinical identification of aphasia, standardized psychometric tests are used which also quantify aphasia severity by assessing fluency (usually subjectively using rating scales), word retrieval, repetition, sentence, discourse production and comprehension [3]. The popular Western Aphasia Battery (WAB) [3] is such a test, which has been adapted in languages other than English. Depending on patterns of performance, the WAB enables clinicians and researchers to classify aphasia into *fluent* (e.g., *anomic, transcortical sensory, Wernicke's*) and *non-fluent* typologies (e.g., *Broca's, transcortical motor*). The WAB also generates severity scores that range from 1 to 100. The lower the score, the greater the severity. Within this range, a score of ≥ 93.8 is thought to be "within normal limits" [3]. This study focuses on people whose WAB scores fall within the normal limits, i.e. latent aphasia.

A score "within normal limits", however, does not mean that a person's language abilities are indeed within normal limits – they can present with latent aphasia on more sensitive measures. [4] compared a group of individuals with latent aphasia with neurotypical controls and found that speech rate, rate of lexical errors and lexical diversity were poorer than controls. Similar findings have been reported in other studies, especially delays in spoken word production [5] and verbal short-term memory difficulties [6]. [7, 8] observed differences in planning sentences and discourse macrostructure which were evident in temporal measures (pause durations, speech and articulation rates). The authors interpreted the longer pause durations and lower speech rate (in comparison to neurotypical controls) as a deficit arising from cognitive processing speed limitations, similar to other researchers [5].

Little is known about the speech and language abilities in latent aphasia and milder aphasia in general. One reason may be that some researchers group such individuals together with other aphasia types and do not study them as discrete entities [4]. Furthermore, the limited research may relate to the diverse labels that have been used to describe such individuals, for example, *subliminal aphasia*, *very mild aphasia,* and *post-stroke cognitive impairment*.

### 1.2. Prosodic research in aphasia

A recent systematic review of prosodic research in aphasia [9] identified prosodic differences between people with aphasia and neurotypical controls, e.g. longer syllable and utterance durations in speech by people with aphasia. Additionally, people with aphasia exhibited difficulties controlling prosodic modulations such as inconsistent f0 in different linguistic contexts. Production issues at the prosody-syntax interface were also evident.

However, most studies that investigated prosodic features in aphasia research used unnatural speech elicitation tasks such as reading single words, phrases, or sentences [9, 10]. Such tasks, while providing tight experimental control, have limited ecological validity. This is because linguistic planning is prescribed by the stimuli and therefore planning demands are lessened in oral reading in comparison to demands in spontaneous speech. Another shortcoming of the prosodic aphasia literature is that it focused predominantly on moderate types of aphasia. To our knowledge, [10] is the only study that focused on mild anomic aphasia and investigated prosody. In comparison to neurotypical controls, the aphasia groups showed greater variability in fundamental frequency (f0) than controls, but the difference was not statistically significant [10]. In summary, there is a need to broaden our understanding of prosodic abilities in even milder types of aphasia than anomic.

### 1.3. Automatic classification of aphasia subtypes

Since people with latent aphasia often perform "within normal limits" in clinical tests and many tests are not sensitive enough to identify them, there is a need to develop more sensitive tools for clinical use to facilitate classification. Correct classification would lead to decisions for treating these subtle problems and therefore provide support to affected persons. Previous efforts have been made for the classification of primary progressive aphasia, caused by dementing conditions rather than stroke [11, 12]. [11] developed auto-classification for primary progressive aphasia with 201 participants' brain-imaging data. The accuracy of the classification reached 92.2%. [12] used features extracted from text transcriptions of storytelling of the Cinderella story, e.g. syntactic complexity measures, part of speech, and word frequency. They trained their models using 40 participants' data (16 controls) and performed Leave-One-Out cross-validation. The results hit as high as 100% accuracy. However, leaving one data point out each time results in the same speaker's data being in both the training and the test set; therefore, the models in this study could hit such a high accuracy rate. The severity of the progressive aphasia in both studies was not mild. Consequently, it remains to be seen whether automatic classification can be achieved in people with very mild or latent aphasia.

### 1.4. The present study

Our study aims to (1) examine how latent aphasia affects the production of prosodic features, especially in different word positions, compared with neurotypical controls, in order to set the foundation for further understanding the role of prosodic features in cognitive-linguistic processing in these individuals. This aim was motivated by recent efforts to relate prosodic features to cognitive-linguistic processes [13]; (2) investigate how prosodic features contribute to the diagnostic classification of people with latent aphasia versus neurotypical controls [11]; and (3) explore whether using all available information, including prosodic features and beyond, we are able to build a fast and reliable tool to assist the classification of people with latent aphasia versus neurotypical controls.

## 2. Data

We used recordings of 20 speakers' retellings of the Cinderella story from AphasiaBank [14]. Ten speakers had latent aphasia (WAB severity mean = 97.2, sd = 1.8) and 10 neurotypical controls. Two-tailed Mann-Whitney tests showed no statistical between-group differences (p < .05) in age, education, and sex.

Annotation of story episodes, utterances, and target words (utterance-initial and utterance-final) was made in Praat [15]. We manually annotated the episodes of the Cinderella stories following a published protocol [16]. Episodes provide the macro-structure of the narrative from beginning to end. For example, utterances that relate to the transformations of Cinderella, mice and pumpkin by the fairy godmother belong to a single episode, different from the episode comprising utterances that convey the prince's search for the foot of the lady who wore the glass slipper. The episodes provide discourse structure and the time course information of the target words, i.e. whether the target words are at the beginning of the story or towards the end of the story. Utterance segmentation was identical to the one in AphasiaBank [14]. All utterance-initial and utterance-final target words were also manually annotated. When target words were affected by noise, the entire utterance was excluded from the final analysis. A total of five utterances were removed from the latent group and four from the control group, leaving 228 and 431 utterances for analysis for the latent and control groups respectively.

## 3. Prosodic differences: regression

First, we investigate whether there are systematic prosodic differences between the utterance-initial and utterance-final words in different speaker groups. The choice of utterance-initial and final words was based on studies in typical [17] and aphasia [18] literature.

### 3.1. Methods

Three features which are the most common acoustic correlates of prosody were extracted using Praat: mean f0 words (extracted with a range of 60 Hz-500 Hz), mean intensity, and duration. Following [18], all three measurements were extracted from utterance-initial and utterance-final words respectively to compare across groups and word positions.

For the analysis, linear mixed effect models were built with *lme4* package [19] in R Studio [20] to model each of the dependent variables (mean f0, mean intensity, mean duration) with fixed effects of SPEAKER GROUP (latent, control), WORD POSITION (initial, final), and their interaction. The full model included a nested structure of EPISODE ID: UTTERANCE and SPEAKER. When a model did not converge, EPISODE ID was excluded.

### 3.2. Results

The mean values and standard deviations of the three prosodic measures are shown in Table 1. Table 2 shows the linear regression results for all three models.

For mean **f0** values, an effect of word position was found: the utterance-final (F) f0 was significantly higher than word-initial (I) f0. The control group produced the utterance-final words with significantly higher mean f0 values than the utterance-initial words, while the aphasia group used significantly higher f0 in utterance-initial words than in utterance-final words. These results suggest that the groups used f0 differently. Similarly, mean **intensity** was significantly higher at utterance-final position than utterance-initial position for the controls, but lower for the latent group. Finally, mean word **durations** were significantly longer for the aphasia group. For both groups, the utterance-final words were significantly longer than the utterance-initial words.

Table 1: *Descriptive results by groups and word positions (I: utterance-initial; F: utterance-final).*

| Group | Position | Mean F0 (Hz) | Mean Intensity (dB) | Mean Duration (ms) |
|---|---|---|---|---|
| control | I | 135.62±37.82 | 68.16±8.29 | 267.84±181.47 |
| | F | 149.6±67.95 | 68.71±7.76 | 424.15±158.1 |
| latent | I | 159.92±44.09 | 62.62±9.92 | 338.5±219.58 |
| | F | 157.51±52.45 | 62.16±10.08 | 506.43±210.1 |

Table 2: *Linear mixed effect models results.*

| | Predictors | Estimates | *t* | *p* |
|---|---|---|---|---|
| **Mean f0** | (Intercept) | *136.28* | *16.56* | *<0.001* |
| | group [latent] | 20.00 | 1.71 | 0.102 |
| | position [F] | **14.42** | **4.13** | **<0.001** |
| | group [latent] × position [F] | **-17.30** | **-2.98** | **0.003** |
| **Mean Intensity** | (Intercept) | 67.93 | 25.50 | <0.001 |
| | group [latent] | -6.33 | -1.68 | 0.110 |
| | position [F] | 0.54 | 1.96 | 0.050 |
| | group [latent] × position [F] | **-1.05** | **-2.22** | **<0.050** |
| **Mean Word Duration** | (Intercept) | 0.27 | 12.92 | <0.001 |
| | group [latent] | **0.07** | **2.44** | **<0.050** |
| | position [F] | **0.16** | **13.10** | **<0.001** |
| | group [latent] × position [F] | 0.01 | 0.53 | 0.597 |

## 4. Classification: random forest models

Understanding the effect of word position on prosodic features in different groups was not the only end of our investigation. We also aimed to understand the role of prosodic features among other variables, including other acoustic features and available demographic information. Therefore, we constructed random forest models due to its high performance for processing high-dimensional data [21] to examine (1) whether using easily acquirable speech and demographic information can reliably predict the classification of the aphasia group and the controls; (2) how important the prosodic features are among all the features. To reliably explore these questions, three different validation methods were used in three different experiments.

### 4.1. Methods

Three feature sets, including two acoustic ones and a demographic information set, were used in the experiments. The first set contained 23 interpretable key acoustic and temporal measurements extracted using a customised Praat script, related to prosody, including, for example, duration, mean intensity, mean f0, minimum and maximum f0 and their time points, jitter, and shimmer. A second set of 988 acoustic features were extracted with the *emobase* feature set in the *openSMILE* [22] Toolkit in Python 3.9. This toolkit was developed for audio feature extraction and classification of speech and music signals and the *emobase* feature set contains features designed for emotion recognition and thus has plenty of prosodic features in the set. A third feature set involved demographic information from [14], i.e., sex, age, and education (number of years).

All models were built in R [20]. *Borutta* package [23] was used for feature selection, and only the confirmed attributes (i.e. the important factors) were included in the final model for training. The *randomForest* package [24] was used for model training. The *caret* package [25] was used for the k-fold cross-validation. All models were trained with complete cases, i.e. when no feature contains any NA values. Different cross-validation methods were used in the three experiments.

### 4.2. Results

Table 3 below shows the results for all three experiments.

Table 3: *Results of all three experiments.*

| | | Accuracy | Sensitivity | Specificity |
|---|---|---|---|---|
| **Exp. 1** | Random split (70% train; 30% test) | 97.8% | 0.986 | 0.971 |
| | 10-fold cross-validation | 99.9% | 0.975 | 0.988 |
| **Exp. 2** | Leave-One-Subject-Out | 74.2% | 0.841 | 0.643 |
| **Exp. 3** | Balanced test set | 78.4% | 0.619 | 1 |

#### 4.2.1. Experiment 1

In this experiment, we aimed to find out whether the available features were sufficient to predict the classification.

Two data validation methods were used: (1) a random 70%-30% train-test split was used, and (2) a 10-fold cross-validation method, i.e. to randomly divide the dataset into ten groups, hold out one group as the test set and train with the rest of the nine groups each time and repeat until all ten groups have been used as the test group, the final mean accuracy is then taken as the final result.

As shown in Table 3, the random split method resulted in a high accuracy of 97.8% in distinguishing the latent aphasia group from controls. The 10-fold cross-validation showed even higher accuracy in the classification – 99.9% on average. The cross-validation methods used in this experiment, especially the 10-fold cross-validation, follow similar principles to the Leave-One-Out cross-validation method used in [12], i.e. to divide the train and test sets randomly; therefore, the results showed as high accuracy as [12].

#### 4.2.2. Experiment 2

While experiment 1 achieved near-perfect performance, it was not the best validation method for evaluating whether a model was robust enough to be used as a classification tool since part of the training set and the test set may contain the same speakers. Therefore, it was essential to evaluate the usefulness of the models again by using the Leave-One-Subject-Out cross-validation method, i.e. holding one speaker out as the test set and using the rest of the speakers as training data. Two speakers, one from the latent aphasia group and one from the control group, were removed from the dataset because of extremely few data points left and missing labels after the cases containing NAs were removed. Test sets only contained complete cases.

As shown in Table 3, the mean accuracy across all predicted data reached 74.2% (sd = 0.364). When counting by speakers, 72.2% of the speakers, 6 speakers with latent aphasia and 7 controls, had more than 50% of correctly classified data points. This is still a satisfactory result, considering the entire dataset only contained 20 speakers and the speakers produced speech ranging from 48 s to 420 s.

#### 4.2.3. Experiment 3

While Experiment 2 provided a fair way of evaluating the models, the training and test datasets were unbalanced. In this experiment, we held out four out of the 20 participants (two from the latent aphasia group, and two from the control group), ensuring each pair of speakers was matched on sex, age range, and the number of tokens produced, as shown in Table 4, so that they were representative of the entire dataset.

Table 4: *Matching demographic information of the test set participants.*

| Group | Age | Sex | Tokens |
|---|---|---|---|
| control | 58 | M | 42 |
| latent | 59 | F | 36 |
| control | 41 | M | 66 |
| latent | 36 | F | 41 |

For these four speakers, as shown in Table 3, the sensitivity (i.e. predicting controls correctly) was not very high, but the specificity (predicting aphasic group correctly) was completely correct. The no information rate, i.e. if all data points are counted in the aphasia group, was 0.568, and the accuracy of 0.784 was significantly higher ($p < 0.001$).

To further investigate the role of prosodic features, we also examined the top ten important factors in the model. The top six and tenth key factors were all intensity-related Pulse Code Modulation; the 7$^{th}$ to the 9$^{th}$ factors were f0-related factors. This shows that prosodic factors were overpoweringly important in predicting the classification of speaker groups.

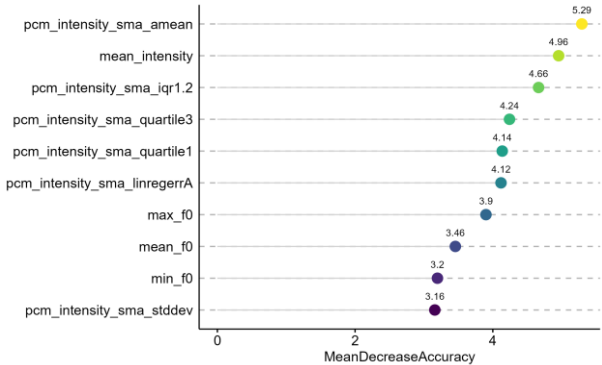

Figure 1: *The top-10 important features in the data for predicting the classification of speaker group. The higher the Mean Decrease Accuracy value, the more important it is in predicting the classification.*

## 5. Discussion

Both the linear regressions and the random forest results suggest that prosodic features are different across the speaker groups and the locations of the target words, and the prosodic features play important roles in predicting classifications.

### 5.1. Linear regression findings

Previous research [7, 8] identified subtle but statistically reliable differences in terms of discourse planning and sentence production in latent aphasia. In the present study, we investigated prosodic variables in utterance-initial and utterance-final words. The two groups differed in mean f0 and intensity in utterance-initial and utterance-final word positions. Following a cognitive approach to explaining prosodic markers [13], our findings could reflect different manners in the way the two groups executed speech plans at the beginning and end of utterances. The longer word durations in the latent group could reflect slower speaking rates or processing speed limitations [5]. In this study, we disregarded the complication of prosodic structure and the lexical and syntactical information of the target words or utterances. Future studies could further analyze text-based features (e.g. syntactic complexity, part-of-speech variables) such as those in [12].

### 5.2. Classification findings

The combination of features extracted with Praat and *openSMILE*, along with the demographic features, were found to be effective and representative of all the features needed for an automatic classification tool. *OpenSMILE* features have achieved good performances in many studies, e.g. in the classification of people suffering from depression [26], and in the classification of autism spectrum disorder [27]. However, we acknowledge that some of the parameters may present strong collinearity. Strong collinearity may result in overfitting of the models; though we did not observe that in our random forest models. The *openSMILE* features were carefully chosen and similar parameters represent the relevant acoustic reality from slightly different perspectives. Moreover, it is not at all an issue for random forest models. However, we do plan to investigate further how to reduce the number of features, both to reduce potential collinearities between factors, and to make the models faster to run and easier to interpret.

On the other hand, we would also like to investigate how to further automate the classification system. Currently, the data were extracted based on utterances which were manually segmented. In our future investigations, all data will be automatically extracted. This will progress towards an automated tool which only makes use of automated feature extraction and model training. Such a tool could produce speech-based classifications which can provide clinicians with a reference for their diagnostic decisions in mild cognitive-linguistic problems in different disorders beyond stroke. Future models could also explore the sensitivity in capturing changes following speech, language, and broader cognitive interventions in different populations.

Lastly, including more person-specific metrics from cognitive tests such as short-term memory and attention, may be able to significantly improve model prediction ability. This is another direction that we plan to explore in future studies.

## 6. Conclusions

To conclude, we conducted prosodic analyses in people with latent aphasia as a way of improving our understanding of this poorly understood type of aphasia and explored the classification potential of machine learning techniques. Despite the limited number of factors we explored in our regression analysis, our findings highlight the merits of prosodic research in identifying subtle pathological differences, paving the way for future research in subclinical and hidden cognitive-linguistic problems. The random forest modelling further supports the regression findings, as the prosodic features demonstrated significant importance. The performance of the random forest modelling also suggests a high likelihood of developing online speech-based assessment tools for clinical uses.

## 7. Acknowledgements

We wish to thank the participants of *AphasiaBank* as well as all its other contributors for making this project possible.